\documentclass{emulateapj}
\usepackage{graphicx}
\usepackage{amssymb}
\usepackage{amsmath}

\begin{document}

\title{Power Spectrum Sensitivity and the Design of Epoch of Reionization Observatories}
\author{Miguel F. Morales}
\affil{MIT Center for Space Research\\77 Massachusetts Ave.\\Cambridge, MA 02139}
\email{mmorales@space.mit.edu}

\begin{abstract}
Recent theoretical developments for observing the Epoch of Reionization (EOR) have concentrated on the power spectrum signature of redshifted 21 cm emission.  These studies have demonstrated the great potential of statistical EOR observations, however, the sensitivity calculations for proposed low frequency radio arrays have been highly approximate.  The formalism developed for interferometric measurements of the cosmic microwave background can be extended to three dimensions to naturally incorporate the line-of-sight information inherent in the EOR signal.  In this paper we demonstrate how to accurately calculate the EOR power spectrum sensitivity of an array, and develop scaling relationships which can be used to guide the design of EOR observatories.  The implications for antenna distribution, antenna size, and correlator requirements on the EOR sensitivity are detailed.
\end{abstract}

\section{Introduction}

In the last year significant advances have been made in understanding the characteristics of the redshifted 21 cm emission from the Epoch of Reionization, and how this signal can be separated from the contaminating foreground emission.  
It was originally believed that contamination from a host of strong foreground sources would overwhelm the EOR signal \citep{DiMatteoForegrounds}, making observations all but impossible.  In the fall of 2003 it was shown by \citet{ZaldariaggaPow1} and \citet{MoralesEOR1} that differences in the frequency characteristics between the foreground contamination and the EOR emission could be used to identify the cosmological signal.  Additionally, these papers show how statistical techniques developed for measurements of the cosmic microwave background (CMB) could be extended to EOR observations.  Recent theoretical advances by \citet{FurlanettoEORPSTheory,FurlanettoEORPSExamples1} have demonstrated that these statistical EOR observations can easily differentiate between various reionization histories. 

The combination of new analysis methods, accurate theoretical calculations, and advances in low frequency radio instrumentation have spawned several experimental EOR efforts.  One remaining stumbling block for instrumental designers is the difficulty in predicting the sensitivity of a radio array to the EOR power spectrum.  In Section \ref{SensitivityDerivation} we generalize the sensitivity calculations to accurately include both frequency information and the effects of antenna distribution.  Sections \ref{ScalingRelationships} and \ref{ConfigurationSec} then explore the effects of design choices on the EOR sensitivity, and the implications for EOR arrays.

It is important to note that residual errors from subtracting foreground sources will contaminate the power spectrum.  The sensitivity calculation presented here only includes the effects of the thermal noise, and thus represents an upper limit on the signal-to-noise ratio which can be obtained by a given array.  However, determining the noise-dominated sensitivity is still important for designing EOR observatories.

\section{Sensitivity Derivation}
\label{SensitivityDerivation}

As discussed in \citet{MoralesEOR1}, the EOR produces a spherically symmetric three dimensional power spectrum signal.  This introduces a notational difficulty since most of the theoretical work for the CMB uses spherical harmonics.  Because the cosmic microwave background signal forms a spherical two dimensional surface on the sky, spherical harmonics are the natural orthogonal set. The $m$ terms of the spherical harmonics can be summed over because the isotropy of space removes the rotational dependence, leaving the standard multipole $l$ expansion of the CMB.  This formalism has trouble with the EOR signal, however, because the observed frequency of the redshifted 21 cm emission maps to the line-of-sight distance, and introduces a third dimension to the data set. One approach is to use a separate spherical harmonic expansion for each ``step'' in the line-of-sight distance, producing a series of shells (see \citet{ZaldariaggaPow1}).  In this representation the signal in neighboring frequency slices is highly correlated due to the large amount of low frequency power in the matter density power spectrum.  This correlation between shells makes calculating the the sensitivity of multi-frequency observations with the spherical harmonic formalism very difficult.

\begin{figure*}
\begin{center}
\plotone{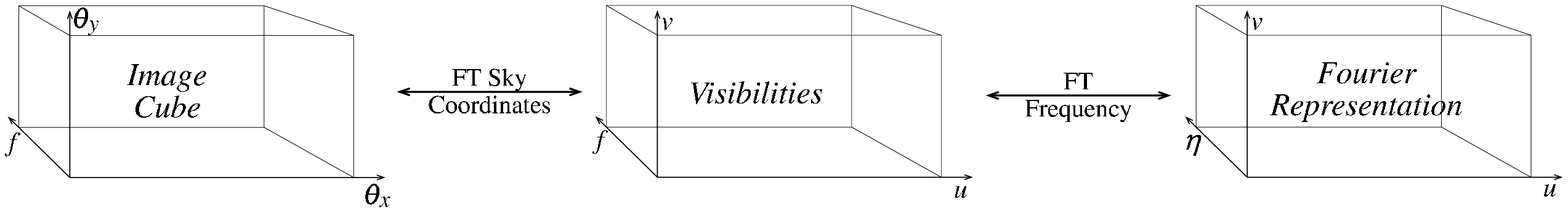}
\caption{This diagram shows the Fourier transform relationships between the image cube, the measured visibilities, and the Fourier representation.  For an interferometer, the fundamental observable is the visibility-frequency cube, which can then be transformed into either an image cube to show source locations (including line-of-sight) or the Fourier representation to analyze the spatial structure of the signal.}
\label{cubeTransform}
\end{center}
\end{figure*}

The solution is to choose a basis set that extracts the spatial structure of the signal in all three dimensions, minimizing the correlation of the parameters.  While there are several basis sets which could be used, we follow the approach of \citet{MoralesEOR1} and use the three dimensional Fourier transform representation.  If spherical sky effects can be ignored, the true signal is uncorrelated in this representation with the only correlation being introduced by the field of view and bandwidth of the observations.  Because the quantities are largely independent in this basis, calculating the sensitivity of an array is much simpler.  The relationships between the image cube, the measured visibility cube, and the Fourier representation are depicted in Figure \ref{cubeTransform}.

The following derivation is based on the visibility calculation developed by \citet{White} for CMB interferrometers, generalized to three dimensional Fourier coordinates.  This approach accounts for the true antenna distribution of the array and the effects of earth rotation when calculating sensitivity of a radio interferometer to the EOR signal, but neglects the effects of foreground contamination. Section \ref{SignalSec} briefly reviews the derivation of the EOR signal power spectrum, and Section \ref{NoiseSec} derives the characteristics of the thermal noise and explains how to precisely calculate the power spectrum sensitivity.  These relationships are then approximated in Sections \ref{ScalingRelationships} and \ref{ConfigurationSec} to explore the effects of array design on the sensitivity.

\subsection{Signal Properties}
\label{SignalSec}
The true distribution of neutral hydrogen emission is given by 
\begin{equation}
\label{trueVis}
\triangle I_{T}(\mathbf{u})=F_{\rm HI}(\mathbf{u}) \equiv C(\mathbf{u}) \,\frac{\rho_{\rm HI}(\mathbf{u})}{\left \langle \rho_{\rm H}\right \rangle}.
\end{equation}
The vector $\mathbf{u}$ describes the position in the three dimensional Fourier space and is defined by $\mathbf{u} \equiv u\hat{\i} + v\hat{\j} + \eta \hat{\rm k}$.  The variables $u,v,\eta$ in turn are the wave numbers associated with the spatial positions on the sky $\theta_{x},\theta_{y}$ and the frequency $f$, as described in Figure \ref{cubeTransform}. (Note that this three dimensional definition of $\mathbf{u}$ is not equal to the two dimensional version used in spherical harmonic CMB derivations.)  

The contributions to the neutral hydrogen emission $\triangle I_{T}(\mathbf{u})$ can be separated into a spin temperature component $C(\mathbf{u})$, and a density and ionization fraction term $\rho_{\rm HI}/\left \langle \rho_{\rm H}\right \rangle$, where $\rho_{\rm H}$  and $\rho_{\rm HI}$ are the density of hydrogen and neutral hydrogen respectively, and $\langle\rangle$ denotes the average.  For a baryon density $\Omega_{\rm B}h^{2} = 0.02$ and a helium fraction $Y = 0.24$, $C$ is given by 
\begin{equation}
\label{ }
C_{\rm K} \equiv (2.9\ {\rm mK})\,h^{-1}\frac{(1+z)^{2}} {E(z)}\frac{\left(T_{\rm s}-T_{\rm CMB} \right)}{T_{s}}
\end{equation}
in brightness temperature units.  
A full derivation of $C$ along with definitions of all terms is given in Appendix A of \citet{MoralesEOR1}. 

The theoretical expectation for the cross-correlation of the true sky emission is a delta function given by
\begin{equation}
\label{ }
\left \langle \triangle I_{T}(\mathbf{u})\,\triangle I_{T}^{\ast}(\mathbf{u}') \right \rangle =  \left \langle |\triangle I_{T}(\mathbf{u}')|^{2} \right \rangle \delta(\mathbf{u}-\mathbf{u}')\,d^{3}\mathbf{u}'.
\end{equation}
However, correlations are introduced by the finite field of view and bandwidth of the experiment.  The correlation between two measured $\triangle \tilde{I}(\mathbf{u})$ values (no subscript $_{T}$ for measured value) is then given by
\begin{multline}
\label{Csignal}
C_{ij}^{I} \equiv \left\langle \triangle \tilde{I}(\mathbf{u}_{i}) \,\triangle \tilde{I}^{\ast}\! (\mathbf{u}_{j})\right\rangle =\\ \int \! P_{\rm HI}(|\mathbf{u}|)\,W(\mathbf{u}_{i}-\mathbf{u})  W^{\ast}\!(\mathbf{u}_{j}-\mathbf{u})\,d\mathbf{u},
\end{multline}
where $W(\mathbf{u}_{i}-\mathbf{u})$ is the window function given by the field of view and bandwidth of the array and $P_{\rm HI}(\mathbf{u})\equiv \left \langle |\triangle I_{T}(\mathbf{u})|^{2} \right \rangle \delta(\mathbf{u}-\mathbf{u}')$ and is spherically symmetric due to the isotropy of space (see \citet{MoralesEOR1} for a full derivation).

Equation \ref{Csignal} captures the essential characteristics of the signal.  The correlation between separate measurements is entirely determined by the field of view and bandwidth of the measurement, and falls rapidly as the measurements move apart in the $\mathbf{u}$ space.  As the field of view and bandwidth increase $W(\mathbf{u}'-\mathbf{u})$ becomes more centrally condensed, reducing the correlation length between measurements and increasing the expectation value of $C_{ij}^{I}$.

\subsection{Thermal Noise Properties} 
\label{NoiseSec}

The fundamental observable for an interferometric array is a single visibility $V$ (Jy).  The thermal noise per visibility\footnote{In Equation \ref{VisNoise}, $V_{\rm rms}$ is implicitly averaged over all of the polarizations observed at a given location (up to 3 for very wide field detectors).  For most ground based arrays there are two polarizations and $T_{\rm sys}$ is reduced by $\sqrt{2}$ from that of a single dipole.  Alternatively, $T_{\rm sys}$ can be defined as the system temperature of a single polarization, and number of independent visibility points in Equation \ref{CnoiseAvg1} can be increased by the number of polarizations.} is given by
 \begin{equation}
\label{VisNoise}
V_{\rm rms} = \frac{2 k_{B}T_{\rm sys}}{A_{e}\sqrt{df\, \tau}},
\end{equation}
where $T_{\rm sys}$ is the total system temperature, $A_{e}$ is the effective area of one antenna (m$^{2}$), $df$ is the bandwidth of a single frequency channel (Hz), and $\tau$ is the total integration time for that visibility measurement (s).

Moving to the Fourier representation, the total noise for every measurement $\triangle I^{N}\!(\mathbf{u})$ can be calculated by Fourier transforming the visibilities. (The superscript $^{N}$ indicates noise and superscript $^{I}$ signal.)  Since the noise contributions to the visibilities are Gaussian distributed with zero mean, $\triangle I^{N}\!(\mathbf{u})$ will also be Gaussian distributed with zero mean.  Since the rms of $\triangle I^{N}\!(\mathbf{u})$ is constant for all $\mathbf{u}$ it can be easily calculated from the $\eta = 0$ term and the average visibility rms
\begin{subequations}
\label{ }
\begin{align}
\label{}
   \triangle \tilde{I}_{\rm rms}^{N}\!(u,v,0)& = \int V_{\rm rms}(u,v,f)\, df,   \\
      &=\sum^{B/df} V_{\rm rms}(u,v,f)\, df,
\end{align}
\end{subequations}
where $B$ is the total bandwidth of the measurement.  The rms of the sum is $\sqrt{N} = \sqrt{B/df}$ times the average rms of $V$ times $df$, and is given by
\begin{equation}
\label{ }
\triangle \tilde{I}_{\rm rms}^{N} = \frac{2 k_{B}T_{sys}\sqrt{B}}{A_{e}\sqrt{\tau}}.
\end{equation}
It is convenient to rewrite the effective area and bandwidth in terms of the physical size of the antenna $dA$ (m$^{2}$) and the inverse of the bandwidth range $d\eta$ (Hz$^{-1}$)
\begin{equation}
\label{Inoise}
\triangle \tilde{I}_{\rm rms}^{N} = \frac{ 2 k_{B}T_{sys}}{\epsilon\, dA\,d\eta\sqrt{B\tau}}.
\end{equation}
In physical terms, $dA\,d\eta$ represents the approximate dimensions of the area averaged by the correlator in the Fourier space, with the efficiency $\epsilon$ representing the collecting efficiency of the array both spatially within an antenna and in frequency. 

To compare with the correlation matrix of the signal $C_{ij}^{I}$ we need to determine the correlation matrix of the noise 
\begin{subequations}
\label{CnoiseCorrA}
\begin{align}
C_{ij}^{N}(\mathbf{u}) &= \left\langle \triangle \tilde{I}^{N}\!(\mathbf{u}_{i}) \,\triangle \tilde{I}^{N\ast}\! (\mathbf{u}_{j})\right\rangle \\
&= \left\langle \left | \triangle \tilde{I}^{N}\!(\mathbf{u}) \right |^{2} \right\rangle \delta_{ij}.
\label{CnoiseCorr}
\end{align}
\end{subequations}
Because the thermal noise is random, $ \langle \triangle \tilde{I}^{N}\!(\mathbf{u})  \rangle$ is Gaussian distributed with zero mean and rms $ \triangle \tilde{I}_{\rm rms}^{N}\!(\mathbf{u})$ given by Equation \ref{Inoise}.  The expected distribution of $ \langle | \triangle \tilde{I}^{N}\!(\mathbf{u}) | \rangle$ is Rayleigh distributed and $ \langle | \triangle \tilde{I}^{N}\!(\mathbf{u}) |^{2} \rangle$ is exponentially distributed with rms $2 [  \triangle \tilde{I}_{\rm rms}^{N}\!(\mathbf{u}) ]^{2}$.  This gives the rms of the noise correlation matrix
\begin{equation}
\label{CnoiseBase}
\left [C_{ij}^{N}(\mathbf{u})\right ]_{\rm rms} = 2\left (\frac{ 2 k_{B}T_{sys}}{\epsilon\, dA\,d\eta}\right )^{2}\frac{1}{B\,\tau}\delta_{ij}.
\end{equation}

Due to sky rotation the values of $\mathbf{u}$ observed by any given pair of antennas are a function of time.  Instead of using $i$ and $j$ to label the observed baselines, equivalently the Fourier space can be divided into a large number of cells with $i$ and $j$ labeling the cells (also applies to Equations \ref{Csignal} and \ref{CnoiseCorrA}). The noise in an analysis cell is then given by
\begin{equation}
\label{CnoiseU}
\left [C_{ij}^{N}(\mathbf{u})\right ]_{\rm rms}= 2\left (\frac{ 2 k_{B}T_{sys}}{\epsilon\, dA\,d\eta}\right )^{2}\frac{1}{B\,\bar{n}(\mathbf{u})\,t}\delta_{ij},
\end{equation}
where $\bar{n}(\mathbf{u})$ is the time average number of baselines in an observing cell and $t$ is the total observation duration. For many realistic arrays  $\bar{n}(\mathbf{u})$ may range from greater than one at short distances due to redundant baselines to much less than one on longer baselines.  The effects of array configuration on the power spectrum sensitivity are explored in Section \ref{ConfigurationSec}.

In principle, Equations \ref{Csignal}, \ref{CnoiseCorr} and \ref{CnoiseU} contain all the information required to calculate the sensitivity of any given array and observing program.  Using a simulation of the detector array and observing strategy, the observed Fourier space can be divided into many small cells, with the signal strength and correlation for the cells given by Equation \ref{Csignal} and the uncertainty in each cell due to thermal noise given by Equation \ref{CnoiseU}.\footnote{Note that the distribution of $C_{ij}^{N}(\mathbf{u})$ for each cell is exponentially distributed, not Gaussian.}  This allows the layout of the array and the observing strategy to be accurately included in the sensitivity calculation.

Equation \ref{CnoiseU} can be further simplified by making use of the spherical symmetry of the signal to average over all of the cells within a spherical annulus.  Except at the shortest baselines, typical arrays have a large number of independent cells in each annulus, and in this limit the mean of the $C_{ij}^{N}(\mathbf{u})$ measurements will become Gaussian distributed with rms
\begin{equation}
\label{CnoiseAvg1}
\left [\overline{C_{ij}^{N}(\mathbf{|u|})}\right ]_{\rm rms} \approx 2\sqrt{\frac{d^{3}u'}{2\pi |\mathbf{u}|^{2}du}}\left (\frac{ 2 k_{B}T_{sys}}{\epsilon\, dA\,d\eta}\right )^{2}\frac{1}{B\,\bar{n}(|\mathbf{u}|)\,t}.
\end{equation}
The $\sqrt{\frac{d^{3}u'}{2\pi |\mathbf{u}|^{2}du}}$ term is simply $N^{-1/2}$ with $N$ being the number of cells within an annulus, with $d^{3}u'$ as the size of the Fourier cells, $du$ the width of the annulus, and the $2\pi$ term being due to the Hermitian nature of the visibility measurements limiting the summation to a half sphere.

Equation \ref{CnoiseAvg1} be used to directly calculate a close approximation of the sensitivity as a function of wave number (one dimensional).  While equations \ref{Csignal}, \ref{CnoiseCorr}, \ref{CnoiseU} and \ref{CnoiseAvg1} allow calculation of the power spectrum sensitivity,  the implications for array design are not immediately obvious.  In the next two sections we explore the scaling behavior of these equations and the implications for array design.

\section{Scaling Relationships}
\label{ScalingRelationships}


Looking closely at Equation \ref{Csignal} for the signal properties reveals two important characteristics.  First, the correlation length is inversely proportional to the field of view and bandwidth of the observation.  For many interferometric arrays at low radio frequencies, the incident electric field is directly sampled by simple dipole or similar detector elements.  Because digitization and correlation is still relatively expensive compared to the cost of a single detection element, it is common to use a beamformer to combine the data from a small group of elements into one ``antenna.''  The field of view of an observatory is roughly equal to the inverse of the physical extent of the beamformed antennas, and we identify $dA$ in our derivation as the area of one beamformed region.  Thus as the size of the antennas shrinks, the field of view expands, allowing a larger portion of the sky to be simultaneously observed and reducing the correlation length of the power spectrum.  

We can use this to go a step farther and choose $dA\,d\eta$ as the cell size for the analysis.  All measurements which fall within one cell will then be highly correlated and measurements in all other cells will be largely independent.  We can then make the approximation that 
\begin{multline}
\label{pkconclusion}
C_{ij}^{I}(\mathbf{u}) \approx \left\langle \big\lvert \triangle \tilde{I}(\mathbf{u})\big \rvert^{2} \right\rangle \delta_{ij} \\ = \int P_{\rm HI}(\mathbf{u}) \, \left\vert W(\mathbf{u}' - \mathbf{u})\right\vert^{2}d^{3}\mathbf{u}'.
\end{multline}
Since the window function $W(\mathbf{u}' - \mathbf{u})$ is sharply peaked, the magnitude of the signal $C_{ij}^{I}$ is proportional to the integral of $|W(\mathbf{u}' - \mathbf{u})|^{2}$.  By Parseval's relation this gives
\begin{equation}
\label{CsignalScaling}
C_{ij}^{I} \propto \frac{\epsilon^{2}}{dA\,d\eta}.
\end{equation}

Rewriting Equation \ref{CnoiseAvg1} with the cell size $dA\,d\eta$ gives us the uncertainty due to thermal noise in our measurement of the EOR power spectrum within an annular shell of width $du$
%
\begin{equation}
\label{CnoiseAvg2}
\left [\overline{C_{ij}^{N}(\mathbf{|u|})}\right ]_{\rm rms} \approx 2\sqrt{\frac{dA\,d\eta}{2\pi |\mathbf{u}|^{2}du}}\left (\frac{ 2 k_{B}T_{sys}}{\epsilon\, dA\,d\eta}\right )^{2}\frac{1}{B\,\bar{n}(|\mathbf{u}|)\,t}.
\end{equation}
In the limit that the noise uncertainty dominates the ``cosmic variance'' uncertainty introduced by the finite number of measurements (S/N $\ll 1$ for each cell), we can divide the expected signal by the rms in an spherical annulus to determine the approximate signal to noise ratio for the EOR power spectrum. We can then vary Equations \ref{CsignalScaling} and \ref{CnoiseAvg2} with respect to different parameters to get the scaling relationships shown in Table \ref{scalingTable}.  As the thermal noise becomes small these scaling laws must be modified, but they hold over the range of sensitivities expected for next generation radio arrays. Table \ref{scalingTable} can help guide experimentalists on how design choices affect the sensitivity of an array to the EOR power spectrum.

\begin{deluxetable*}{rcccccc}
  \tablehead{Equation (\#) & $A_{t}|_{dA}$ & $A_{t}|_{N_{\!A}}$ & $dA|_{A_{t}}$ & $B$ & $|\mathbf{u}|$ & $t$  }
  \startdata
  \vspace{.1 in}
  $C_{ij}^{I}$ (\ref{CsignalScaling}) & --- & $A_{t}^{-1}$& $(dA)^{-1}$& $B$ & --- & ---    \\
   \vspace{.1 in}
  $\triangle \tilde{I}^{N}_{\rm rms}$ (\ref{Inoise}) & --- & $A_{t}^{-1}$ & --- & $B^{1/2}$ & --- &$t^{-1/2}$   \\
  \vspace{.1 in}
$\left [\overline{C_{ij}^{N}(\mathbf{|u|})}\right ]_{\rm rms}$ (\ref{CnoiseAvg2}) & $A_{t}^{-2}$ &$A_{t}^{-5/2}$& $(dA)^{-1/2}$ & $B^{1/2}$ & $[|\mathbf{u}|\,\bar{n}(|\mathbf{u}|)]^{-1}$ & $t^{-1}$   \\
  \vspace{.1 in}
Power Spectrum S/N & $A_{t}^{2}$ & $A_{t}^{3/2}$ & $(dA)^{-1/2} \propto$  FOV & $B^{1/2}$ &  $|\mathbf{u}|\,\bar{n}(|\mathbf{u}|)$ & $t$  
\enddata
\tablecomments{This table lists the scaling relationships of the key equations.  In order, the variables in each column are: total array area holding the size of each antenna constant $A_{t}|_{dA}$ (adding antennas), total array area holding the number of antennas and distribution constant $A_{t}|_{N_{\!A}}$ (increasing antenna size), the size of each antenna with the total array area held constant $dA|_{A_{t}}$ (dividing area into more small antennas), the total bandwidth $B$, the sensitivity as a function of wavenumber length $|\mathbf{u}|$, and the total observing time $t$.}
\label{scalingTable}
\end{deluxetable*}


One of the unusual properties of the power spectrum measurement is that it is fundamentally the square of an intensity, so the uncertainty of the power spectrum decreases as the square of the uncertainty in the individual measurements.  This leads to surprising effects, such as the uncertainty being inversely proportional to the integration time $t$ instead of $\sqrt{t}$---a well known if seldom publicized effect from CMB experiements.  However, not all scaling relations are squared.  Certain parameters such as the bandwidth do not decrease the uncertainty of an individual power measurement but instead add independent power measurements, and the sensitivity scales with the square root of the number of measurements.  To help work through all of the various combinations in the table, the following list explains the behavior observed in each column of the table.
\begin{itemize}
  \item {\bf Adding Antennas $\left( A_{t}|_{dA} \right )$.} Adding collecting area by adding more antennas of the same size ($dA$ held constant) decreases the uncertainty in the power spectrum measurement as $A_{t}^{-2}$.  This dependence comes entirely through the average number of measurements per cell $\bar{n}(|\mathbf{u}|)$.  Because the size of the antennas is constant the cell size remains the same while the number of visibilities increases as the square of the number of antennas, increasing $\bar{n}(|\mathbf{u}|) \propto A_{t}^{2}$.  Of course the correlation load also increases proportional to $A_{t}^{2}$, which can be a significant expense.
  \item {\bf Increasing Antenna Size $\left( A_{t}|_{N_{A}} \right )$.} Another alternative for adding collecting area is to increase the size of each antenna while leaving the number and distribution of antennas the same.  The advantage of this scheme is that the correlation load is unchanged, however, the gain in sensitivity is only proportional to $A_{t}^{3/2}$ because the field of view of the array is reduced, effectively decreasing the number of independent measurements.  Mathematically this can be seen by scaling with respect to $dA$ with $\bar{n}(|\mathbf{u}|) \propto dA$ since the average number of measurements per cell increases with the cell size.
  \item {\bf Fewer Elements per Antenna $\left(dA|_{A_{t}}\right )$.} An alternative to adding collecting area is to simply increase the number of antennas by putting fewer detection elements into each antenna group.  This increases the correlation load, but the sensitivity also increases with $dA^{-1/2}$ because the field of view increases.  Mathematically, vary with respect to $dA$ again, but this time $\bar{n}(|\mathbf{u}|) \propto dA^{-1}$ because the total number of visibility measurements scales as $dA^{-2}$ but the average increases linearly with the cell size.
  \item {\bf Increasing Bandwidth $\left(B\right )$.} Increasing the bandwidth adds entirely new measurements without increasing the sensitivity of the array to the line emission of individual H$_{\rm I}$ regions, so the sensitivity scales with the number of new measurements as $B^{1/2}$.  Mathematically, $d\eta \propto B^{-1}$ and the density of measurements per cell $\bar{n}(|\mathbf{u}|)$ is independent of the bandwidth (cell size decreases with increase in number of frequency channels).
  \item {\bf Sensitivity as a Function of Wavenumber $\left(|\mathbf{u}|\right )$.} As one moves to larger wavenumbers (smaller length scales), the number of cells increases as $|\mathbf{u}|^{2}$ leading to a linear dependence of the sensitivity on $|\mathbf{u}|$.  Of course this is very sensitive to the density of measurements at that wavenumber $\bar{n}(|\mathbf{u}|)$ as discussed in Section \ref{ConfigurationSec}.
  \item {\bf Increasing Integration Time $\left(t\right )$.} Increasing the total observation time decreases the uncertainty on each measurement, leading to a linear increase in sensitivity with duration.  This somewhat counterintuitive result is because the power spectrum is related to the square of the intensity.
\end{itemize}

It should be remembered that these scaling relationships are only valid when the signal-to-noise ratio is small in each cell.  \citet{HalversonThesis} gives a nice description of the transition from noise to statistics dominated uncertainty and the scaling within the noise dominated regime for CMB interferometers.  These scaling relationships can help array designers determine the effects of their design choices, and are useful for comparing the sensitivity of different EOR arrays.  In the next section the effects of antenna placement are further explored.

\section{Effects of Array Configuration}
\label{ConfigurationSec}

One of the unusual properties of interferometric arrays is that the distribution of visibility measurements is easily adjusted by altering the layout of the antennas.  This freedom allows one to fine tune the characteristics of the array to maximize the scientific return.  In this section we explore the effects of antenna layout on the EOR power spectrum sensitivity.

The quantity $\bar{n}(|\mathbf{u}|)$ represents the average number of measurements per Fourier cell in a spherical annulus of radius $|\mathbf{u}|$.  This can be calculated from average number of measurements in each individual cell $\bar{n}(\mathbf{u})$ by averaging over all the cells at that annulus
\begin{equation}
\label{NtrueAvg}
\bar{n}(|\mathbf{u}|) = \frac{\int_{0}^{\pi}\!\!\int_{0}^{\pi} \bar{n}(\mathbf{u})\, |\mathbf{u}|^{2}\sin\theta\,d\phi\, d\theta}{\int_{0}^{\pi}\!\!\int_{0}^{\pi} |\mathbf{u}|^{2}\sin\theta\,d\phi\, d\theta},
\end{equation}
where $\theta$ is measured from the $\eta$ axis, and $\phi$ is the azimuthal angle and ranges from $0 \rightarrow \pi$ since we only integrate over half the sphere because the visibilities are Hermitian. 

\begin{figure}
\begin{center}
\plotone{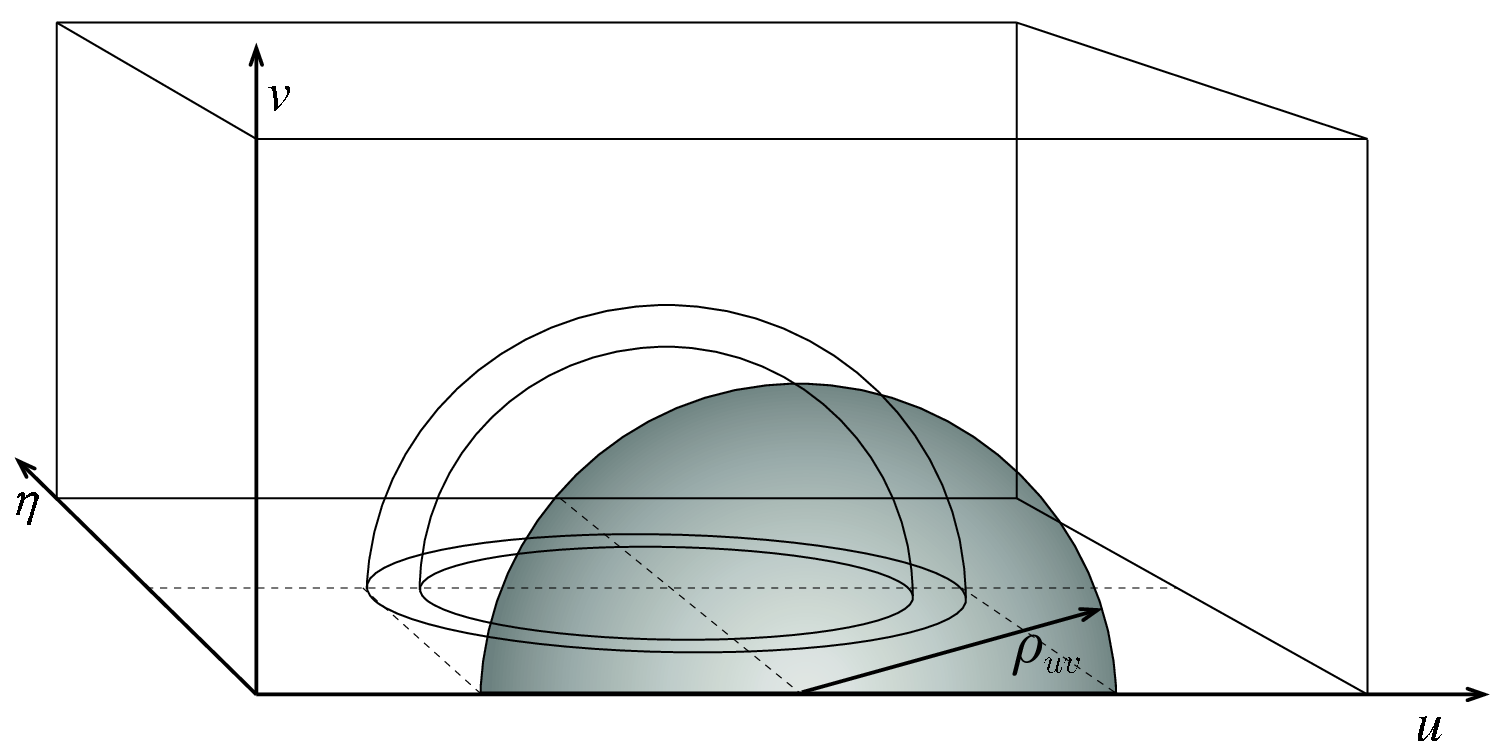}
\caption{The measurement of the power spectrum at a scale $|\mathbf{u}|$ is performed by combining all of the measurements within a spherical annulus at that diameter, as shown above (the origin is at the center of the annulus where the dashed lines cross).  The shaded region shows the contribtuion of antenna baselines in the $u,v$ plane to this measurement, with dark indicating a higher contribution.  All baselines $\rho_{uv} \leq |\mathbf{u}|$ contribute to the measurement with a limb brightening functional distribution. Even very short baselines contribute information at high $|\mathbf{u}|$ because of the spatial information encoded in the frequency dimension.  The presence of the third dimension fundamentally changes the characteristics of the measurement and makes EOR experiments less sensitive to the antenna distribution than similar CMB interferometers.}
\label{distSensitivity}
\end{center}
\end{figure}

This relationship can be simplified by concentrating on the spatial distribution of the visibility measurements.  If we neglect the small distortion introduced by fractional bandwidth, the density of measurements per cell is constant for all cells at a position $u, v$, independent of $\eta$.  This is because each baseline measures all frequency values, thus the density of measurements is just a function of $u$ and $v$.  If we use polar coordinates $\rho_{uv}$ and $\phi$ to represent the position of a baseline in the $u,v$ plane (see Figure \ref{distSensitivity}), Equation \ref{NtrueAvg} can be simplified to 
\begin{equation}
\label{ }
\bar{n}(|\mathbf{u}|) \approx \frac{1}{\pi |\mathbf{u}|}   \int_{0}^{\pi}\!\!\int_{0}^{|\mathbf{u}|}\bar{n}(\phi, \rho_{uv}) \frac{\rho_{uv}}{\sqrt{|\mathbf{u}|^{2}-\rho_{uv}^{2}}}\,d\phi\,d\rho_{uv},
\end{equation}
after carrying out the integral in the denominator and changing variables.  
$\bar{n}(\phi, \rho_{uv})$ is simply the average number of baselines within a two dimensional cell in the $u,v$ plane during the course of the observation. For many experiments the $u,v$ coverage is very nearly azimuthally symmetric, allowing one to simplify even further to obtain
\begin{equation}
\label{NsimpleAvg}
\bar{n}(|\mathbf{u}|) \approx \frac{1}{|\mathbf{u}|}   \int_{0}^{|\mathbf{u}|}\bar{n}( \rho_{uv})\frac{\rho_{uv}}{\sqrt{|\mathbf{u}|^{2}-\rho_{uv}^{2}}}\,d\rho_{uv}.
\end{equation}

Under closer inspection, Equation \ref{NsimpleAvg} provides a very interesting result:  the sensitivity of the array at a given scale $|\mathbf{u}|$ depends on all the visibility measurements with a spacing equal to or less than this scale.  The very shortest baselines still contribute to measurements normally associated with much larger antenna spacings.  The reason for this effect is the three dimensional distribution of the measurements shown in Figure \ref{distSensitivity}.  Because spatial information related to the line-of-sight distance is encoded in the observed frequency, very short baselines still contain information about very large $|\mathbf{u}|$ at the appropriate $\eta$ value. The weighting of the baseline contributions has the same pattern as limb brightening. This means that the sensitivity of an EOR interferometer as a function of scale is fundamentally different than a CMB interferometer.  Some of the sensitivity distributions which have appeared in the literature implicitly assume a single frequency of observation (2D) and do not accurately represent the scale dependence of proposed measurements.  

\begin{figure}
\begin{center}
\includegraphics[height = 6.5 in.]{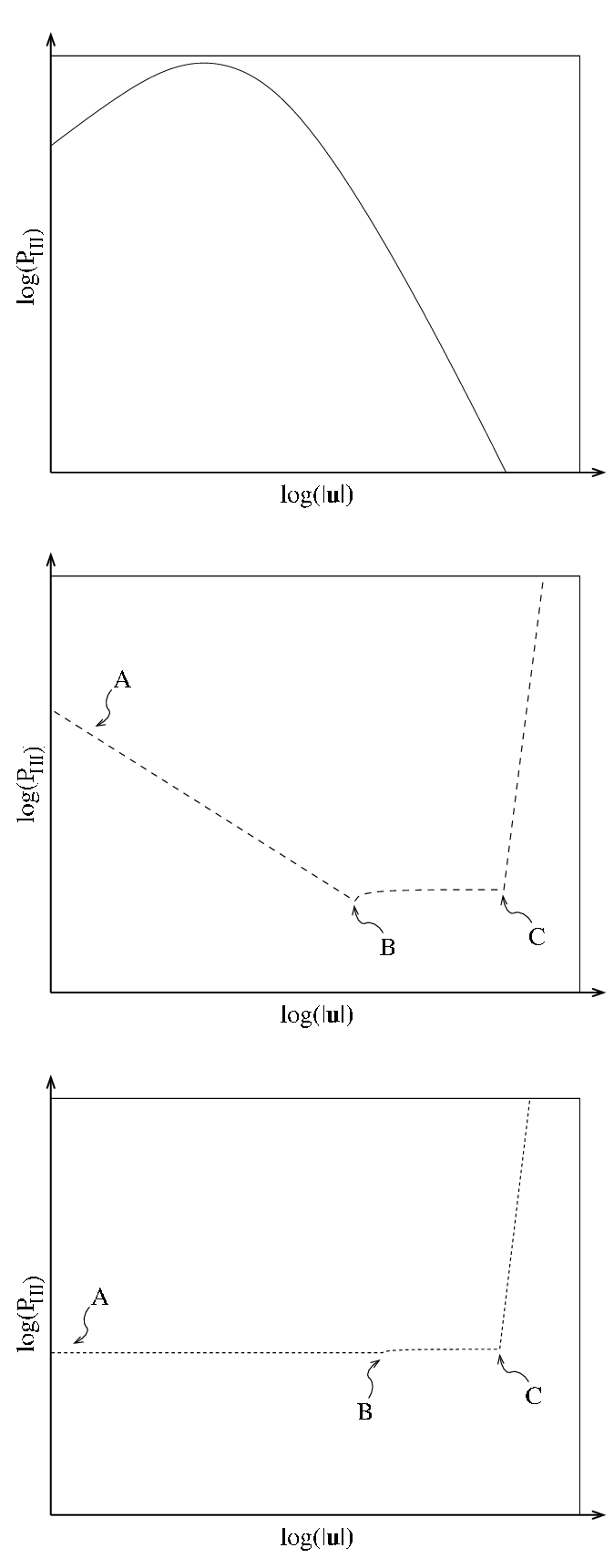}
\caption{This series of cartoons shows the shape of the power spectrum sensitivity as a function of $|\mathbf{u}|$ for different antenna distributions.  Because the normalization and characteristics of the sensitivity depend strongly on the specific features of a particular array, no scale is presented on these graphs.  Instead we are concentrating on the shape of the distributions.  The top panel shows the power spectrum signal for a hot fully neutral medium.  The middle and bottom panels show the uncertainty due to thermal noise (noise power spectrum, for sensitivity in bins of constant logarithmic size multiply by $|\mathbf{u}|^{-1}$) .  The middle panel has the distribution for a uniformly sampled $u,v$ plane, and the bottom panel  has a  $u_{\rho}^{-1}$ distribution similar to that found in centrally condensed distributions.  At the small baselines indicated by marker A, the limited number of measurements becomes important, increasing the uncertainty above that indicated by the thermal noise curve (similar to ``cosmic variance'' effects). Additionally, for centrally condensed arrays, the $u_{\rho}^{-1}$ visibility distribution can only be maintained until neighboring antennas touch, leading to additional modifications at short baselines.  Point B indicates the largest baseline of the array, and point C the length scale associated with the frequency channel width.  Typically C is at a larger $|\mathbf{u}|$ than B, and the sensitivity extends to length scales normally associated with larger baselines.  After point C the number of observations quickly goes to zero. }
\label{sensitivityWScale}
\end{center}
\end{figure}

Figure \ref{sensitivityWScale} shows the scale dependence of the sensitivity for two example antenna distributions.  The uniform $u,v$ coverage distribution is very difficult to realize in practice, but does approximate the $u,v$ distribution generated by a ring of antennas and has been used as a simplifying assumption in the literature.  The other distribution is for a centrally condensed array with a $\bar{n}(\rho_{uv}) \propto \rho_{uv}^{-1}$ distribution, approximately the distribution given by a centrally condensed antenna distribution.  The highest spatial frequency which can be observed in the $\eta$ dimension is set by the inverse of the frequency channel resolution, and is determined by the design of the correlator.  For most interferometers narrow channel widths lead to the $\eta$ dimension of the observation extending to much larger values than the $u$ and $v$ dimensions.  This leads to edge effects where the annulus becomes larger than the longest baseline and then larger than the $\eta$ extent (see Figure \ref{sensitivityWScale}).  

The ideal antenna configuration depends both on the scales of $|\mathbf{u}|$ which provide the best differentiation between models and the need to minimize systematics.  Neglecting systematics favors very compact arrays since short baselines will still contribute to large $|\mathbf{u}|$ measurements through the frequency domain.  However, realistic EOR interferometers are all expected to include longer baselines.  These longer baselines are crucial for identifying and removing a host of systematic effects, such as the point source foreground, ionospheric refraction, radio interference and radio transients.  Determining the optimum array layout will require an accurate model of the expected signal, the contaminating effects, and the analysis method, but will likely result in a centrally condensed array with a diameter of a couple of kilometers.

\section{Conclusion}

This paper has developed the formalism needed to calculate the power spectrum sensitivity of multi-wavelength EOR observations and has explored the implications for array design.  By examining the scaling relationships and the effects of antenna distribution on the sensitivity, design trade-offs can be balanced against one another.  It is hoped that this work will help optimize the design of low frequency radio telescopes for EOR observations.


\section*{Acknowledgments}
I would like to thank Colin Lonsdale, Roger Cappallo, and Jackie Hewitt for carefully reading and correcting early drafts.  I would also like to thank the entire Mileura Wide-field Array team for blazing the path towards building a premier EOR observatory in Western Australia.  This work has been supported by NSF grant AST0121164.

 \end{document}